\documentclass[conference]{IEEEtran}
\IEEEoverridecommandlockouts
\usepackage{cite}
\usepackage{amsmath,amssymb,amsfonts}
\usepackage{algorithmic}
\usepackage{graphicx}
\usepackage{textcomp}
\usepackage{xcolor}
\usepackage{float}
\usepackage{tabularx}
\usepackage{graphicx} 
\usepackage{bm}
\usepackage{dsfont}
\usepackage[caption=false]{subfig}
\usepackage{xstring} 
\usepackage{booktabs}
\usepackage{comment}
\usepackage{hyperref}
\begin{document}

\title{Community Formation and Detection \\ on GitHub Collaboration Networks\\
}
\author{\IEEEauthorblockN{Behnaz Moradi-Jamei}
\IEEEauthorblockA{\textit{School of Data Science } \\
\textit{ University of Virginia }\\
Charlottesville, United States \\
}
\IEEEauthorblockA{\textit{ Department of Mathematics \& Statistics} \\
\textit{ James Madison University}\\
Harrisonburg, United States \\
 moradibx@jmu.edu}
\and
\IEEEauthorblockN{Brandon L. Kramer}
\IEEEauthorblockA{\textit{ Biocomplexity Institute} \\
\textit{University of Virginia}\\
Arlington, United States \\
kb7hp@virginia.edu}
\and
\IEEEauthorblockN{J. Bayoán Santiago Calderón}
\IEEEauthorblockA{\textit{Biocomplexity Institute} \\
\textit{University of Virginia}\\
Arlington, United States \\
jbs3hp@virginia.edu}
\and
\IEEEauthorblockN{Gizem Korkmaz}
\IEEEauthorblockA{\textit{Biocomplexity Institute} \\
\textit{University of Virginia}\\
Arlington, United States \\
gk8yj@virginia.edu}

}

\IEEEaftertitletext{\vspace{-2\baselineskip}}

\maketitle
\vspace{-5mm}
\begin{abstract}

This paper studies community formation in OSS collaboration networks. While most current work examines the emergence of small-scale OSS projects, our approach draws on a large-scale historical dataset of 1.8 million GitHub users and their repository contributions. OSS collaborations are characterized by small groups of users that work closely together, leading to the presence of communities defined by short cycles in the underlying network structure. To understand the impact of this phenomenon, we apply a pre-processing step that accounts for the cyclic network structure by using Renewal-Nonbacktracking Random Walks (RNBRW) and the strength of pairwise collaborations before implementing the Louvain method to identify communities within the network. Equipping Louvain with RNBRW and the contribution strength provides a more assertive approach for detecting small-scale teams and reveals nontrivial differences in community detection such as users' tendencies toward preferential attachment to more established collaboration communities. Using this method, we also identify key factors that affect community formation, including the effect of users' location and primary programming language, which was determined using a comparative method of contribution activities. Overall, this paper offers several promising methodological insights for both open-source software experts and network scholars interested in studying team formation.
\vspace{-0.5mm}


\end{abstract}


\section{Introduction}
\vspace{-1mm}
Open-source software (OSS) development is characterized by a high level of collaboration from developers worldwide~\cite{octoverse-2020}. Recent scholarship has shed light on the tremendous social and economic impacts of these collaborations. For example, Lima et al.~\cite{lima2014coding}, and other social network analysts find that OSS collaboration systems can develop along with several trajectories but typically exhibit power-law-like distributions of highly active users clustering in hubs of intense development activity. From these hubs, OSS collaboration networks tend to grow into core-periphery structures where small, closely-linked teams drive the project with larger, loosely-coupled users reporting problems or contributing to documentation at the periphery~\cite{crowston2006core, Ducheneaut05socializationin}. While most existing work examines network-level tendencies in smaller-scale projects, the current paper draws on a dataset of 1.8M users scraped from GitHub - the world’s largest remote hosting platform - to examine community formation in large-scale collaboration networks.

More specifically, this paper aims to better explicate how communities are shaped by the cyclic structure of the network rather than just existing edges of a graph. To do this, we introduce a novel pre-processing method for detecting communities that depends on pre-processing the strength of collaborations among users and then incorporating information about the topological structure of the network into our clustering methods. In this approach, we start by pre-processing the network data using Renewal-Nonbacktracking Random Walks (RNBRW)\cite{MORADIJAMEI2021125116} and collaboration strength (CSRBRW) before applying the Louvain community detection algorithm~\cite{blondel2008fast}. This method provides a stronger approach for detecting small-scale team formation by accounting for preferential attachment to more established collaboration communities. Additionally, we identify key factors that impact the formation of communities, including user location and primary coding language. The idea is that users who collaborate with the same coding languages are more likely to collaborate than those with different languages. In order to infer primary coding language for each user, we introduce a procedure that compares four assignment rules based on repository-level contribution activity and compares the results on community formation across the network. By combining this procedure alongside the CSRBRW approach, this paper offers several innovative methodological insights that may aid OSS scholars and network scientists interested in studying group formation. 
\subsection {Contributions}
\vspace{-1.5mm}
The main contributions of this paper are threefold: \\
    \emph{1)} We propose a novel pre-processing method for detecting OSS communities by utilizing cyclic properties of the network in combination with the ``collaboration quality" as represented by the network's edge weights. After applying this pre-processing step to calculate contribution scores, we use the Louvain clustering algorithm to detect communities based on network modularity. Our proposed method reveals the importance of cyclic connections in projects that have multiple contributors. \\
    \emph{2)} We develop a hierarchical approach to study how repository- and user-level attributes contribute to community formation. Our main focus in this section is exploring the influence that users' country affiliations and primary programming languages have both across and within communities. While we find that primary language does help explain some of our findings, we observe notable heterogeneity in country representation across these groups, suggesting that location does not do well in explaining community formation tendencies. \\
    \emph{3)} We implement and compare four different rules to assign programming languages to developers. While GitHub users self-report their locations as user-level attributes, primary languages must be inferred from contribution activities to repositories (or repos). We assign the users a primary language based on the four following conditions: (\emph{i}) the language of the repos that they own, (\emph{ii}) the majority language among all the repos they contribute to, (\emph{iii}) the language of the repo they contribute the highest number of commits, and (\emph{iv}) the number of bytes contributed in a given language. We observe that for the majority rule, the distribution of the proportion of the languages assigned to the users is significantly different than those with other rules. 

\vspace{-1mm}
\section{Related Work}
\vspace{-1mm}
Community detection is a prominent topic of study in network science~\cite{fortunato2010community}. Existing methods in this domain typically define community membership based on the most prevalent edges shared between users of a given network. Community detection methods usually aim to optimize a network property like modularity \cite{blondel2008fast}. Alternatively, model-based approaches, such as the stochastic block modeling, discover communities by maximizing the likelihood of realizing an observed graph.~\cite{karrer2011stochastic} While citation counts of prominent papers in the field suggest that modularity-based approaches are more common~\cite{blondel2008fast, clauset2004finding}, a recent study demonstrates strong equivalence between the two approaches~\cite{newman2016equivalence}. Due to the $\mathcal{NP}$-hard nature of optimal community detection, heuristics have been offered for efficiently detecting communities in large-scale network structures with the Louvain algorithm being one of the most popular approaches used to study weighted networks~\cite{blondel2008fast}. Several studies now show that the performance of community detection methods can be enhanced by integrating the structural properties of networks into these algorithms, including node centrality~\cite{khadivi2011network}, common neighborhood ratios~\cite{zhang2015novel}, edge weights~\cite{de2013enhancing}, and the network's cyclic structure~\cite{shakeri2017network}. Unfortunately, due to the computational burdens of implementing such methods, these advances are rarely applied to large-scale networks. Below, we draw on an approach proposed by Moradi et al.~\cite{MORADIJAMEI2021125116} to account for the cyclic structure of networks in order to improve the quality of our community detection method.  
This strategy provides a computationally efficient tool for classifying cycles within networks, which can be scaled for larger networks like those analyzed in this paper. Given that small cycles of collaboration are typical in the open-source software (OSS) community, we expect this method to more appropriately determine which members are relevant to open-source communities - a point that we elaborate in further detail in the coming pages.
 
 


To date, there have been several studies that focus on community detection in real-world social networks, including Facebook~\cite{tamburri2019exploring, 8022716}, Twitter~\cite{huberman2008social}, and Wikipedia~\cite{joblin2017evolutionary}. For example, Win et al.~\cite{8022716} identify communities and outliers on Facebook by relying on degree centrality to determine user similarity and neighborhood overlap. Similarly, Sung et al.~\cite{sung2018uncovering} use a game-theoretic approach to identify overlapping communities, finding that the use of dominance ratios and correlations can reveal how university affiliation, degree major, and temporal effects contribute to stronger group formation on Facebook friendship networks. Jokar and Mosleh~\cite{JOKAR2019718} propose a label propagation based method for community detection in social networks. After applying a pre-processing step that assigns edge weights based on a density score, their method shows the capacity to accurately identify community membership while lowering the computational complexity and time required to do so.  


While several existing studies use network analysis to study OSS collaboration systems, the majority of these papers focus more on user-level tendencies~\cite{cheng2017developer, cheng2019activity} and network-level properties rather than community formation ~\cite{joblin2017evolutionary,tamburri2019exploring}. The few studies that do investigate community structure mostly center around the ``onion’’ model ~\cite{nakakoji2002evolution}, which is a conceptual framework that suggests OSS communities are structured hierarchically with a tightly-knit group members driving the community's evolution at the core and disconnected users at the periphery contributing primarily to debugging and documentation. Network scholars tend to support this view of OSS when they identify core-periphery structures in small-scale network evolution~\cite{crowston2006core, joblin2017classifying, crowston2017core}, but these insights are rarely expanded into community detection on network data. Studies that do attempt to classify communities rely on alternative methods, including commit metrics \cite{DIBELLA201372} and, more recently, decision trees~\cite{agrawal2016resource}. Of course, other factors can contribute to community formation, including various forms of social diversity. Lima et al.~\cite{lima2014coding} have analyzed international collaboration tendencies in GitHub, finding power-law-like distributions form as networks evolve. Moreover, Vasilescu et al.~\cite{vasilescu2015gender} find that gender and tenure diversity contribute positively to team productivity in GitHub collaboration networks. In this paper, we aim to show the advantages of using the CSRNBW approach for community detection in large-scale OSS collaboration networks. 

\section{Data and Methods}
\vspace{-1mm}
\subsection{GitHub Data}
\vspace{-1mm}
In this paper, we use data scraped from GitHub - the world's largest remote hosting code platform. While GitHub allows code hosting for all project types, our main focus is the universe of OSS. To collect OSS projects, we limited our data collection to repositories with one of the 29 machine-detectable licenses approved by the Open Source Initiative (OSI)~\cite{OSI,RubyLicense}. By limiting our analysis to repositories with OSI-approved licenses, we mostly circumvent issues with informal repos like personal projects and homework assignments, opting to analyze development activity that is more likely to be reused. After defining the OSS universe based on this criteria, we used the GHOST.jl~\cite{ghost.jl} package to collect GitHub activity data for all non-forked, non-mirrored, or non-archived (i.e., original content) spanning from GitHub's creation in 2008 through the end of 2019. This \emph{commits table} includes the repository name, user login, date committed, lines added, and lines deleted for all commits made to OSI-licensed repos during this period. The original commits table consists of $\approx$3.26M distinct contributors and $\approx$7.76M distinct repositories. 

\subsection{Network Construction}
\vspace{-1mm}
Using this commits table, we constructed the OSS collaboration networks where nodes correspond to contributors and ties represent the number of shared repositories between two users. If two contributors commit to the same repository, a tie will link these users in the network and the weight of this edge increases as the two contributors jointly commit to more repositories. In our forthcoming analyses, we examine two OSS collaboration networks. In the \emph{\textbf{full OSS network}}, we included all users that collaborate with other users on at least one repository, but removed all of the isolates from the network (i.e., users that only contribute to single-user repositories). In total, the full network includes $\approx$1.8 million nodes and $\approx$147 million edges. Second, we took a subset of the full data based on users that have valid country codes, referred to as \emph{\textbf{international OSS network}}. These country codes are derived from GHTorrent~\cite{gousios2012ghtorrent} -- another large-scale data collection project of GitHub data that provides user attribute data for $\approx$2.1 million users. The \emph{GHTorrent table} includes login names along with self-reported cities, states, countries, and organizational affiliations. To supplement this table, we also developed a script to acquire users' emails. After removing users with missing information, standardizing data from the location columns into country codes, and then joining this data to our original network data, we were left with $\approx$700,000 nodes and $\approx$32 million edges in the international network data. 

Past research has identified automated users (or bots) as a major component of the open-source ecosystem, performing a number of critical maintenance tasks across prominent OSS projects on GitHub~\cite{wessel2018power, golzadeh2021ground}. To address the impact of bots in the full GitHub contributor network, we developed a SQL query that filtered 3,337 users ending in ``bot". We then constructed the contributor network without these bot accounts and compared the network to our full network that included bots, looking specifically to see the difference in the number of communities detected. While only about 1\% of nodes in our networks are bots, we found that those users typically had some of the highest degree centrality rankings in the network, leading us to further examine their impact on community formation. Overall, the largest community in the full network is comprised of $\approx$16\% of all users while the non-bot network's largest community contains about $\approx$15.5\% of all users. The Jaccard similarity on community size between these networks was $\approx$40\%. Considering the difference between the number of nodes in these two networks, the graphs are quite comparable. For this reason, we include bots in our forthcoming analyses. 

\subsection{Community Detection}
\vspace{-1mm}
Prior to performing community detection, we implemented a novel method that analyzes the network's mesoscopic structure using \textit{RNBRW}~\cite{MORADIJAMEI2021125116}. RNBRW is a variant of a random walk where the walk is not only prohibited from returning back to a node in exactly two steps, but also terminates and restarts once it completes a cycle. RNBRW quantifies the cyclic structure by pre-weighting the edges based on this retracing probability. With this approach, the retracing probability $\pi_{e}$ of an edge $e$ (i.e., the likelihood that the edge participates in a cycle) is approximately proportional to the number of times each edge has been retraced by a nonbacktracking random walk. After determining these retracing probabilities for the entirety of the network, the edges are then weighted based on RNBRW weights $\pi_{e}$ multiplied by the number of shared repositories between each of the two collaborators $SC_e$, which we refer to as the Strong Collaborative RNBRW (or CSRNBRW). This can be formalized as $\pi_{e} \times SC_e$.  

This approach is highly scalable and efficient due to RNBRW's parallelizability and effectiveness in capturing network cyclic structure while maintaining the computational advantages achieved by Monte Carlo methods. Moreover, the method performs well on networks with power-law-like features (such as our GitHub networks) because unlike other random walk methods, RNBRW accounts for high degree dependence by keeping the random walk lengths smaller by means of its stopping rule. Table~\ref{louv} shows the performance boosting advantages of RNBRW equipped with the Louvain algorithm to detect communities across various simulated networks with different sizes and average degrees, but the present study would be the first time this strategy is applied to real-world social networks. 

\begin{table}
	\caption{Performance measured by Normalized Mutual Information for Louvain with and without RNBRW weighting for sparse simulated networks with $\hat{d} \approx \log n$. Table adapted from \cite{MORADIJAMEI2021125116}.}
	\vspace{-2mm}
	\resizebox{\columnwidth}{!}{
\begin{tabular}{ |c|c|c| c| }
	\hline
	network size (n) & average degree ($\hat{d}$)  & Louvain & RNBRW+Louvain\\
	\hline
	$10,000$ & $\log n$ & $0.740$   &   $0.970$\\
	& $2\log n$  & $0.757$    &   $\textbf{1}$\\
	& $3\log n$   & $0.882$   &      $\textbf{1}$\\
	\hline	
	$100,000$& $\log n$ &  $0.524$    &   $0.960$\\	
	& $2\log n$ &   $0.649$   &     $\textbf{1}$\\	
	& $3\log n$  & $0.713$   &      $\textbf{1}$\\	
	\hline			  
	$1,000,000$ & $\log n$  & $0.192$   &      $0.969$\\	
	\hline
\end{tabular}}
\label{louv}
	\vspace{-3mm}
\end{table}

\begin{figure}
\vspace{-1.5mm}
	\centering
	{\includegraphics[clip, width=.5\columnwidth]{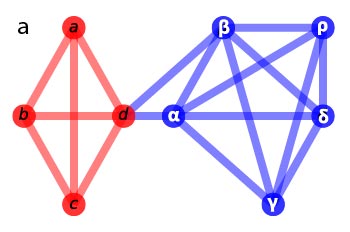}}{\includegraphics[clip, width=.5\columnwidth]{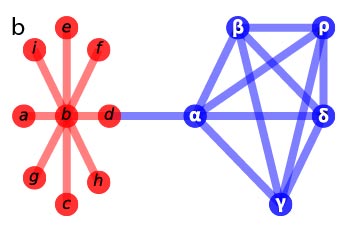}} 
	\vspace{-6mm}
	\caption{Toy networks exemplifying the difference between community detection using Louvain alone compared to the CSRNBRW pre-processing step used alongside of Louvain. Both networks are comprised of users (nodes) with ties that represent collaborations between shared OSS repos. In Figure~\ref{fig:RetracedNBRW}a, the red and blue groupings would be defined as two separate communities using both Louvain and CSRNBRW. In contrast, Figure~\ref{fig:RetracedNBRW}b shows the results when using CSRNBRW to pre-process this graph based on the retracing probabilities. While Louvain alone would define this subgraph as two distinct communities, the pre-processing step would lead to the connected component (in blue) being defined as one community with the star structure (in red) being classified as nine singletons rather than one coherent community. This method emphasizes the importance of network cycles rather than disconnected components like the star structure seen here.}
	\label{fig:RetracedNBRW}
	\vspace{-6mm}
\end{figure}

In essence, the CSRNBRW method is a \textit{pre-processing step} that assigns weights to a network's edges based on their retracing probability times the number of shared repos before applying a community detection method. Figures~\ref{fig:RetracedNBRW}a and \ref{fig:RetracedNBRW}b illustrate how accounting for cycle formation in toy graphs can impact community detection results. Figure~\ref{fig:RetracedNBRW}a displays one set of users $a$, $b$, $c$, $d$ who contribute to a common repository (forming a clique), another set of users $\alpha$, $\beta$, $\gamma$, $\delta$, and $\rho$ that contribute to a second common repo, and a third set of users $\alpha$, $\beta$, and $d$ who contribute to third common repo. This toy graph demonstrates how if more than two users contribute to a shared repo, the collaboration will form a topological cycle in the graph, which will be factored into the community structure. In this case, the CSRNBRW method would identify the two subgraphs in red and blue as two distinct communities just as the Louvain method alone would. 

In contrast, Figure \ref{fig:RetracedNBRW}b provides an example that demonstrates how implementing CSRNBRW as a pre-processing step equipped with Louvain can lead to differences in community detection. In this toy network, users $a$, $d$, $c$, $e$, $f$, $g$, $h$, and $i$ each collaborate with user $b$ in different projects, and users $\alpha$, $\beta$, $\gamma$, $\delta$, and $\rho$ collaborate together on another shared project. While the subgraph on the left forms a star structure, the subgraph on the right forms a cyclic structure. If implemented without the pre-processing step, the Louvain method would detect two communities: $0:\{a, b, c, e, f, g, h,i, d\}$ and $1:\{\alpha, \beta, \gamma, \delta, \rho\}$. However, after the pre-processing step, Louvain only classifies one community: $0:\{\alpha, \beta, \gamma, \delta, \rho\}$, leaving the rest of users classified as their own singleton communities. Together, these toy networks show the performance difference of the Louvain method before and after using the CSRNBRW method.

\subsection{Investigating Community Formation}
\vspace{-1mm}
Once we identified communities using CSRNBRW, we studied how repo- and user-level attributes influence community formation. Currently, node-level data is somewhat difficult to obtain for GitHub data, as this data is only available in the self-reported fields provided on GitHub's site. Since most contributors do not voluntarily (or honestly) fill in these fields, we focus solely on the more widely available features: users' country affiliations and programming languages. 

\subsubsection{Country Affiliations}

To allocate users into countries, we used the \emph{GHTorrent user table} that includes login names along with self-reported cities, states, countries, organizational affiliations, and emails. Because these data are self-reported (and thus quite messy), we developed a series of algorithms to classify users into countries by probabilistically matching users to countries based the location, organization, and email columns. Users that reported being in multiple countries were removed from our international network for simplicity. 

\subsubsection{Language Assignment Rules}
While locations were incorporated by using self-reported user data, inferring programming languages required an additional step to assign this attribute to individual users. To infer users' primary coding language we used \emph{GHTorrent's repository language table}\cite{gousios2012ghtorrent}, which includes the repository name, language contributed, and the number of bytes in that language. To start, we limited our analyses to repos that have a single language. We then joined this table to our \emph{commits table}, resulting in a table with details on the users, the repo they contributed to, the lines of code added/deleted, the language, and the total number of bytes in each language. Since developers can contribute in many languages and the data is not directly connected to users contributions, we developed a set of heuristics to convert these data into user-level attributes. The four rules tested in this paper include: 

\begin{itemize}

    \item \textbf{Rule 1 (Bytes)}: We assign users a primary language based on the language of the repository they contribute to the most as measured by quantity of bytes;
    
    \item \textbf{Rule 2 (Commits)}: We assign languages to users based on the highest number of commits in a given language;  
    
    \item  \textbf{Rule 3 (Majority Language)}: We assign languages to users based on the most common majority language in the repos that they contribute; 
    
    \item  \textbf{Rule 4 (Repo Ownership)}: We assign languages to users based on the most common majority language in the repos that they own (with all other contributors to that repo being assigned primary languages based on bytes as per Rule 1).  
    
\end{itemize}

For more details on our code and methodology, see our \textcolor{blue}{\href{https://github.com/uva-bi-sdad/oss-2020}{GitHub repository}}. 

\vspace{-2mm}
\section{Results}
\vspace{-1mm}
In our results section, we start by comparing community detection methods by contrasting both full and international OSS networks. In these comparisons, we utilize a basic application of the Louvain method and contrast that with Louvain after the CSRNBRW pre-processing step. As we outlined above, community detection methods can rely on several network properties, including the graph's modularity, number of paths, or cyclic structure. The Louvain method relies the modularity of the overall network to detect communities and then draws on each community's modularity to determine whether additional partitions need to be made based on whether that measure is maximized~\cite{blondel2008fast}. Louvain method involves two stages, in the first stage initializes each node into its own community then considers gain in modularity locally by changing a node’s partition label to nodes of a neighboring partition. While this procedure is widely used for assessing the quality of detected communities, modularity-based methods can lead to bias. For example, networks that are partitioned based on modularity may have issues decomposing nodes into the ``correct" communities, leading to over- or under-estimation of the number of communities across the entirety of the network~\cite{fortunato2007resolution}. To account for this potential bias, we report and compare the (a) robustness to resolution limit, (b) the number of communities detected, (c) the community size distribution, and (d) the community network's behavior. In the final section, we extend our analysis to investigate the underlying attributes that may affect community formation, focusing on attributes such as users' country affiliation and their primary programming language. Below, we compare our four assignment rules and investigate the role of these attributes in community detection and formation.

\begin{table}[htbp]
\vspace{-3mm}
\caption{Summary statistics of \\ full and international GitHub networks.}
\vspace{-5mm}
\begin{center}
\resizebox{\columnwidth}{!}{
	\begin{tabular}{|c  |c| c |}
	\hline
	{\bfseries GitHub Network Descriptive} & {\bfseries Full network}& {\bfseries International Network} \\ \hline
        Nodes    & $1.8$M & $459$K \\ \hline
        Edges    & $147$M & $30.8$M\\ \hline
        Density & $0.0009$ & $0.0002$ \\ \hline
        Transitivity & 0.822 & $0.492$ \\ \hline
        Average Degree   & 160 & 134 \\ \hline
        Largest connected component  & $\approx80\%$ nodes & $\approx93\%$ nodes\\ \hline
  
     \end{tabular}}  \\
   \end{center}
   \label{descriptive}
   \vspace{-5mm}
\end{table}

\subsection{GitHub Networks}
\vspace{-1mm}
Table \ref{descriptive} details several metrics for our two GitHub networks, including the node and edge count, density, transitivity, average degree centrality, and the size of the largest connected component. The full GitHub network has $\approx$1.8M nodes and $\approx$147M edges whereas the international network has $\approx$459k nodes and $\approx$30.8M edges. Given that isolates would have no impact on community formation, we excluded those nodes from our analyses. The number of connected components obtained for the full network is equal to 146,652 and the largest
component contains $\approx$80\% of nodes whereas the international network's largest component contains $\approx$93\% of nodes. Density refers to the collaborations between contributors which is defined as the number of edges  a contributor has divided by the total possible connections a contributor could have. As the table shows, both networks have small, relatively comparable densities. Lastly, transitivity refers to the probability that GitHub network have adjacent contributors interconnected which is obtained by the ratio between the observed number of closed collaboration triplets and the maximum possible number of closed collaboration triplets in the network. Table \ref{descriptive} shows that the GitHub full network exhibits considerably higher transitivity than the international network.

\subsection{Number of Communities Detected}

In the full network, the use of Louvain alone identified $\approx$140,000 communities and CSRNBRW+Louvain classified $\approx$40,000 communities. In the international OSS network, Louvain alone identified $\approx$16,000 communities while the application of CSRNBRW+Louvain detected $\approx$13,000 communities. We argue that the drop in the number of communities can be explained by emphasizing cycles and discounting the star structures discussed above (see Figure~\ref{fig:RetracedNBRW}b). We further examined the distribution of community sizes to validate this finding. 

\subsection{Distribution of Community Sizes}
\begin{figure}[t]
\vspace{-1mm}
	\centering
	{\includegraphics[trim= 0 -50 0 10, clip,height=5.5cm,width=0.9\columnwidth]{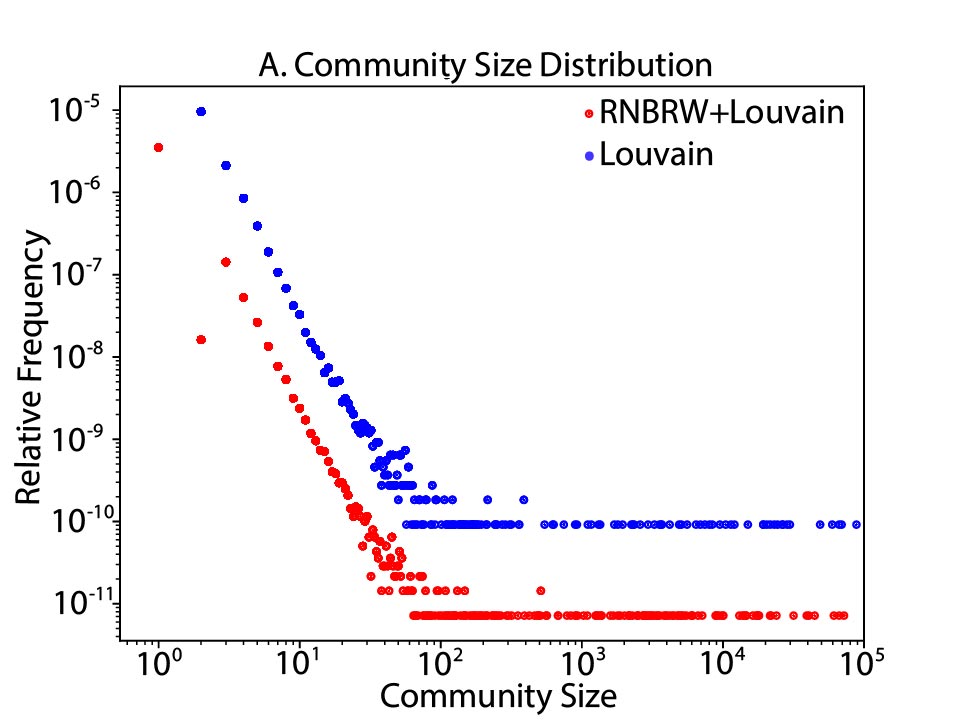}}\\ \vspace{-0.4cm}
	{\includegraphics[trim= 0 0 0 10,clip,height=5.3cm,width=0.9\columnwidth]{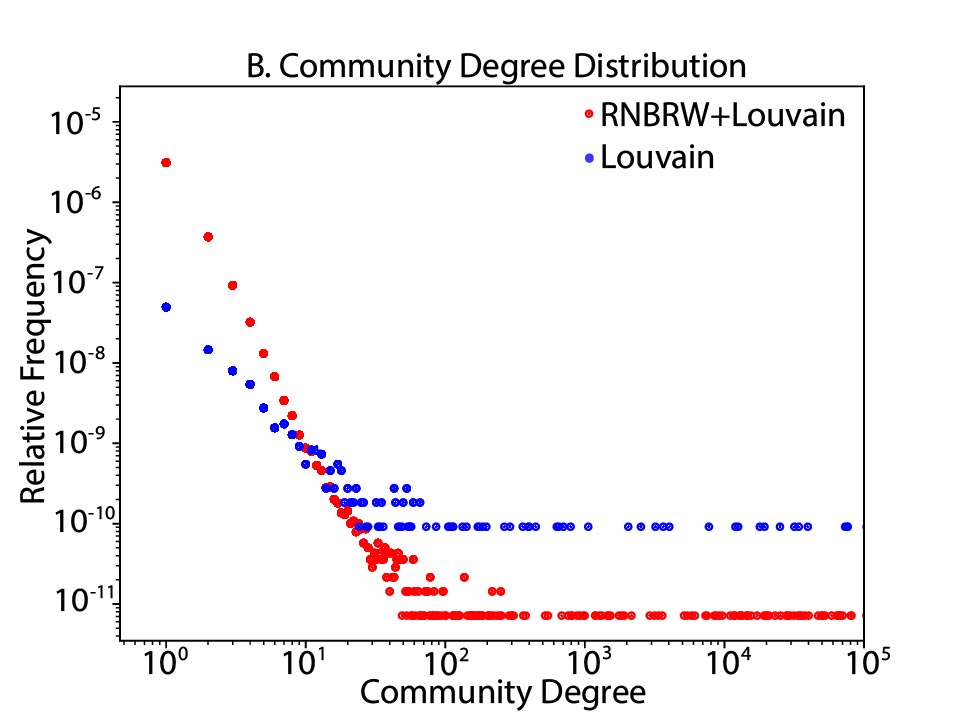}} \\
	\caption{\textbf{Full Network.} (a) The distribution of the community size and \\(b) community degree distribution for the full network on a log-log scale.}
	\label{fig:comm_dist_full} 
\vspace{-5mm}
\end{figure}

\vspace{-1mm}
To analyze the differences in detected communities, we broke down the distribution of the community sizes for both the full and international networks. Figures 2 and 3 show that the community size distributions have similar forms, exhibiting power-law-like distributions with slightly different exponents. Figure~\ref{fig:comm_dist_full}a shows the distribution of full network community sizes that were identified by Louvain compared to the CSRNBRW+Louvain approach. Both methods discovered a similar number of small and small-to-medium-sized communities - ranging from $4$ to $1,000$ members. On the other hand, Louvain alone identified $\approx$105,000 two-member communities whereas CSRNBRW+Louvain found only about $2,000$ two-member communities, showing rather clearly how the pre-processing method ignores dyads in the classification process. When looking at the distribution of communities between $1,000$ and $5,000$ nodes, CSRNBRW+Louvain detected $20$ more communities than using Louvain alone. On the other hand, Louvain revealed $3$ more very large communities of size $50,000$+. In fact, the largest community classified by CSRNBRW+Louvain was $\approx$150,000 where Louvain alone found $3$ communities that exceeded that size with one of these communities being larger than $300,000$. While Louvain (as well as its more recent variant Leiden~\cite{traag2019louvain}) allocates smaller communities into very large communities, it seems that the CSRNBRW+Louvain method breaks these larger communities into sizes of $\approx$1,000 to $\approx$5,000 instead. 

\begin{figure}[]
	\centering
	{\includegraphics[trim= 0 -40 0 10, clip,height=5.2cm,width=0.9\columnwidth]{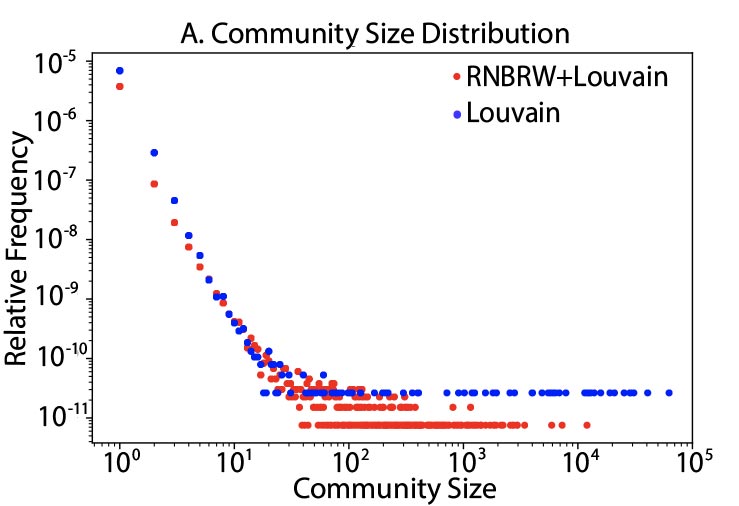}} \\ \vspace{-0.4cm}
	{\includegraphics[trim= 0 0 50 10, clip,height=5.2cm,width=0.9\columnwidth]{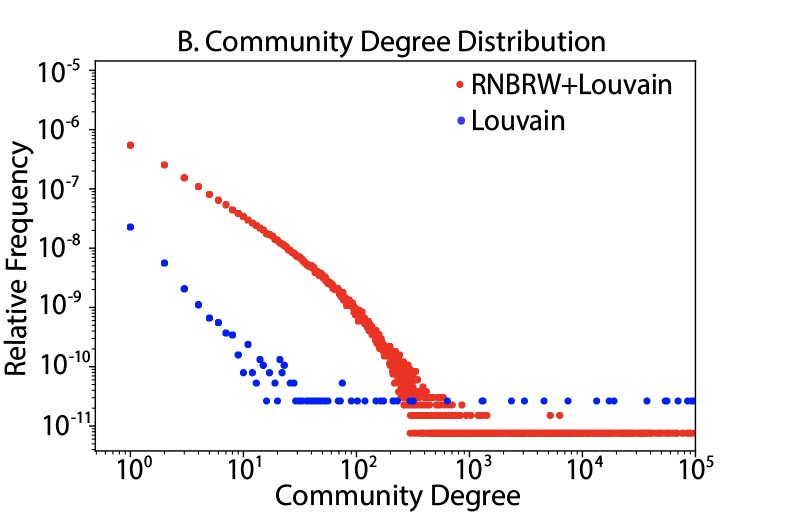}}
	\caption{\textbf{International Network.} (a) The distribution of the community size and (b) community degree distribution for the international network on a log-log scale.}
	\label{fig:comm_dist_intl} 
\vspace{-5mm}
\end{figure}

This is important for network scholars to consider when determining what is practically meant by a ``community." Conceptually, the result of the Louvain-only approach means that $\approx$16\% of the overall network is allocated into the largest community, creating a scenario where the definition of a ``community'' is so broad that it becomes almost impossible that everyone in that group knows or collaborates with each other in any meaningful way. In a recent paper, Wagenseller et al.~\cite{wagenseller2018size} argue that community size should be critical to network scientists' understanding of community detection. Drawing on what anthropologists call the ``Dunbar number,'' these scholars suggest that a community's size should not exceed $\approx$150 due to the cognitive and temporal constraints that individuals have. Following this logic, one of the strengths of the CSRNBRW+Louvain method is that $\approx$94\% of the network falls into group sizes between $3$ and $150$ while the Louvain-only method has only about $\approx$30\% of the network falling into this range. 

Finally, Figure~\ref{fig:comm_dist_intl}a shows the distribution of international network community sizes that are discovered by Louvain and CSRNBRW+Louvain strategies. Both methods discovered approximately 30 communities with a size of $1,000$+ nodes. Louvain classified 25 of these communities into clusters of more than $3,000$ while all of these were under $3,000$ using the CSRNBRW+Louvain approach. Moreover, CSRNBRW+Louvain did a better job of identifying smaller-sized communities of 3-4 people ($\approx$ 1,324) and of 5+ members ($\approx$ 547) than by just using Louvain alone. These findings reflect what past OSS studies identify as the most representative patterns of collaboration in this domain~\cite{crowston2006core, joblin2017classifying, crowston2017core}.

\subsection{Degree Distribution of Community Network}
\vspace{-1mm}
To learn more network structure on collaboration influences, we constructed a \textit{community network} as a weighted graph where nodes correspond to communities and edge weights correspond to the number of collaborations between their members. Figures \ref{fig:comm_dist_full}b and \ref{fig:comm_dist_intl}b provide a graphical demonstration of the number of collaborations among teams that follow a power-laws distributions up to a cutoff (with exponents of 2.8 and 2.3 respectively as obtained using the Louvain+CSRNBRW method). Similar exponents were also found for the international networks. Our observations show that the majority of OSS collaboration networks grow in a scale-free manner that may suggest preferential attachment strategy for the nodes with small/new teams joining existing collaborations with members of more established teams. However, there are few large niches with nodes that possess extreme cumulative advantage that violate this pattern. 

\subsection{Resolution Limit}
\vspace{-1mm}
Modularity maximization methods have limitations for identifying communities with total edges less than $ \sqrt{|E| \times 2}$~\cite{fortunato2007resolution} and with a node count less than $\sqrt{|E|/2}$~\cite{lima2014coding}.
Because of the large number of smaller communities in OSS research, this resolution limit has been shown to pose a challenge to the context we are researching as well~\cite{lima2014coding}. To overcome this limitation, CSRNBRW assigns smaller weights to the edges with a lower probability of merging using Louvain's greedy heuristics. Using just the Louvain method revealed $15$ total communities that went beyond the resolution limit of $\sqrt{|E|/2}$ while the Louvain+CSRNBRW method identified only one community at that threshold for the international network. This corresponds to $0.11\%$ compared to $0.005\%$ of detected communities. Given these findings, we argue that there is a notable advantage to using Louvain+CSRNBRW.



\subsection{ Comparison of Language Assignment Rules}
\vspace{-1mm}
Next, we examine the effect of users' primary language on community formation. This process is challenging because the languages are provided at the repo-level rather than for individual users, which means we need to develop a set of procedures for determining users' primary language. Around $70\%$ of the GitHub repos are assigned a single programming language. Thus, to reduce noise in our assignment strategies, we focus on this subset of the data to test our allocation heuristics of languages to individual users. 

As detailed above, we then tested four different assignment procedures and the impact that they had on which communities were assigned. To evaluate the effect of four assignment rules, we first examined if the distribution of proportion of languages across these assignment rules were significantly different. Table \ref{tab1} reflects the proportion of several top programming languages across the four assignment rules. We observe that the proportion of the languages assigned to the users is very similar across these four rules.

To compare the distribution of number of distinct languages of a community across these rules, we first tested the normality assumption by visually looking at the histograms and distribution of languages across these rules which were highly right-skewed. Because of this non-normal distribution, we performed a  Wilcoxon signed-rank test, which is robust to violation of non-normality. The Wilcoxon signed-rank tests the null hypothesis that two matched samples originate from the same distribution. Before Bonferroni adjustment, each of the $4$ hypotheses were rejected using the significance level of $0.1$. We chose a more conservative $\alpha = 0.05$ level to reduce the possibility of type-2 errors (true-positives) and to increase the power of our tests. Using this threshold, we did not find enough evidence to reject the hypothesis that there is no difference between Rules $1$ (bytes) and $2$ (commits) (\textit{p} = $0.85$) or the hypothesis that there is no difference between Rules $1$ (bytes) and $4$ (ownership) (\textit{p} = $0.73$). On the other hand, three of the \textit{p}-values provide much stronger evidence towards rejecting the hypothesis: pairwise comparisons between Rules $3$ (majority language) and $4$ (ownership), Rules $2$ (commits) and $3$ (majority language), Rules $1$ (bytes) and $3$ (majority language). We conclude that we have strong evidence that distribution of languages in Rule $3$ (majority language) differs from the other assignment rules.


\begin{table}[htbp]
\vspace{-2mm}
\caption{The proportion of top programming languages \\ using each assignment rule}
\vspace{-7mm}
\begin{center}
\resizebox{\columnwidth}{!}{
\begin{tabular}{|p{2cm}|p{1.2cm}|p{1.2cm}|p{1.2cm}|p{1.2cm}|}
\hline
{\bfseries Language} &{\bfseries Rule $1$ } & {\bfseries Rule $2$ }& {\bfseries Rule $3$ }& {\bfseries Rule $4$} \\
\hline
Python &  $0.142$ & $0.150$& $0.145$& $0.153$\\
Java &  $0.103$ &  $0.107$&$0.105$&$0.115$\\
Javascript & $0.169$ & $0.138$&$0.172$&$0.137$\\
Etag&$0.089$ & $0.114$&$0.073$& $0.121$ \\
Php&$0.066$ & $0.063$&$0.067$& $0.065$ \\
C$\#$&$0.043$&$	0.046$&$	0.044$&	$0.048$\\
C$++$&$0.018$ & $0.021$&$0.018$& $0.020$\\
Shell&$0.049$ & $0.047$&$0.049$& $0.040$\\
Ruby &$0.077$ & $0.680$&$0.078$& $0.070$\\
\hline
\end{tabular}}
\label{tab1}
\vspace{-5mm}
\end{center}
\end{table}

\subsection{Community Formation} 
\begin{figure}[htbp]
	\centering
	\label{fig:rule1_70}{\includegraphics[trim= 0 0 10 0,clip,width=.5\linewidth]{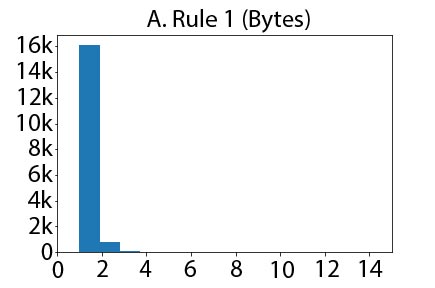}}\label{fig:fig:rule1_all}{\includegraphics[trim= 0 0 20 0,clip,width=.5\linewidth]{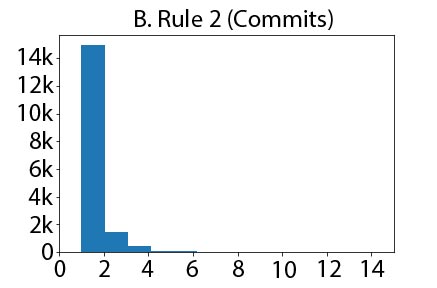}}\\

	\label{fig:rule3_70}{\includegraphics[clip,width=.25\textwidth]{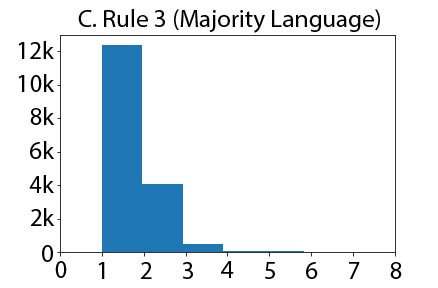}}\label{fig:rule4_70}{\includegraphics[clip,width=.25\textwidth]{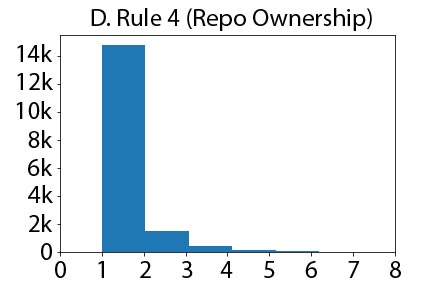}}
	\caption{
	The histograms of number of distinct  languages a community has under four different language assignment rules.
	}
	\label{fig:rules} 
\vspace{-3mm}
\end{figure}

\vspace{-1mm}
In this final section, we investigate the effect of repo and user attributes in community formation. Figure \ref{fig:rules} illustrates the number of communities with distinct languages under the four different assignment rules. This figure shows that Rule 1 allocates over $90\%$ of the detected communities involve contributors with the same language. In contrast, Rule 2, Rule 3, and Rule 4 classify only around $70\%$ of the communities into one programming language. Hence, a considerably large portion of the communities involve GitHub developers contributing with a common community-dependent coding language, suggesting that users' coding languages do play a critical role in forming topological communities.  

The second attribute we investigated was the contributor's country affiliation. We observed considerable membership variation across countries and then subsequently examined country homogeneity across communities with common coding language. To do this, we performed a ${\chi}^2$ test of homogeneity to see if the membership varies based on user's country of origin for the top-10 coding languages. We observe non-significant differences across all languages tested. While rejection of the null hypothesis was expected, we believe that considerably small \textit{p}-values observed come are likely the result of most countries having relatively small overall totals. Future work will likely need to consider continents (e.g., North America, Europe) instead of countries or attempt to over-sample users from less well-represented countries. 

\vspace{-2mm}
\section{Conclusion}
\vspace{-2mm}
In this paper, we analyzed community formation in two large-scale OSS networks using data scraped from GitHub. We implemented a pre-processing stage for OSS community detection that incorporates the networks' cyclic properties and collaboration quality - an approach we call the CSRNBRW method. By comparing the Louvain to the CSRNBRW+Louvain methods in both the full and international GitHub networks, we find marked improvements when comparing various metrics, including the robustness to resolution limit, the number of communities detected, the community size distribution, and the community network's degree distribution. More specifically, we find that the CSRNBRW+Louvain provides more robustness to the resolution limit and detects more desirable sized communities. Furthermore, by investigating the distribution of the OSS team sizes, we observed the tendency of contributors to form small and small-to-medium sized groups, which we argue is a more appropriate approach than allocating users to communities of 150+ people like the Louvain-only method tends to do. Our study also compares four rules for assigning primary languages to users and the impact that this has on community formation. We show that these strategies result in relatively similar results apart from Rule 3 (majority language). Finally, we show that while primary coding language does do well in helping to explain community formation, users' country affiliation does not contribute to these groupings in a meaningful way. 

\subsection{Impact and Future Work}
\vspace{-1mm}
Overall, this work offers insights that both network scholars and OSS developers can benefit. Specifically, we believe that the methodological insights of community detection and the empirical results showing that coding language shapes collaboration tendencies to be important contributions. For example, industry professionals may use our methodology and results to organize their own communities better. Moreover, companies may find this approach useful for recruiting job candidates and understanding how communities work together in and outside of the company's ecosystem. While we believe this is a strong step in the right direction, future work will likely involve defining a more sophisticated approach for assigning primary language based on byte usage. We also plan to consider other user attributes (e.g., repo topics, organizations) to explain community formation and to implement other methodological procedures such as classification trees to predict how these characteristics predict community membership. 

\section{Acknowledgments} 

\vspace{-2mm}

\footnotesize This material is based on work supported by U.S. Department of Agriculture (58-3AEU-7-0074) and the National Science Foundation (Contract \#49100420C0015). The authors acknowledge Research Computing at the Univ. of Virginia for providing computational resources and technical support that have contributed to the results reported within this publication. The views expressed in this paper are those of the authors and not necessarily those of their respective institutions.
\vspace{-2mm}

\bibliographystyle{IEEEtran}
\vspace{-2mm}
\bibliography{bibliography.bib}
\vspace{-2mm}
\end{document}